\documentstyle[12pt,aaspp4]{article}

\def\O{\Omega}
\def\teff{T_{eff}}
\def\cm{{\rm\ cm}} 
\def\cm2g{{\rm ~cm^2~g^{-1}}}
\def\gcm3{{\rm ~g~cm^{-3}}}
\def\ergcm3{{\rm ~ergs~cm^{-3}}}
\def\ecs{{\rm ~ergs~cm^{-2}~s^{-1}}}
\def\g{{\rm ~g}}
\def\s{{\rm ~s}}

\def\pers{~{\rm s}^{-1}} 

\def\ergs{{\rm ergs~s^{-1}}}

\def\msun{{\,\rm M_{\odot}}}

\def\lsun{{\,\rm L_{\odot}}}
\def\msyr{{\msun}~{\rm yr}^{-1}}
\def\lacc{L_{acc}}
\def\linc{L_{inc}}
\def\O{\Omega}

\def\ok{\Omega_K}
\def\oms{\Omega_{_*}} 

\def\omks{\Omega_K(R_{_*})}
\def\omk2i{\Omega_K^2(R_{in})}
\def\omk2o{\Omega_K^2(R_{out})}
\def\omk2s{\Omega_K^2(R_{_*})}
\def\md{\dot M}
\def\ms{M_{_*}}
\def\rs{R_{_*}}

\def\ro{R_{out}}
\def\hs{H_{_*}}
\def\hrs{\hs/\rs}

\def\et{{\it et al.}\ }

\def\sles{\lower2pt\hbox{$\buildrel {\scriptstyle <}
   \over {\scriptstyle\sim}$}}
\def\sgreat{\lower2pt\hbox{$\buildrel {\scriptstyle >}
   \over {\scriptstyle\sim}$}}

\begin{document}

\title{Heating of a Star by Disk Accretion}
\author{Robert Popham}
\affil{Harvard-Smithsonian Center for Astrophysics}
\authoraddr{MS 51, 60 Garden St., Cambridge, MA 02138}

\begin{abstract}

We examine various ways in which disk accretion can heat an accreting
star. These include 1) radiation emitted from the disk surface which
is intercepted by the stellar surface, 2) radiative flux directly
across the disk--star interface, and 3) advection of thermal energy
from the disk into the star. For each of these, the physics of the
boundary layer between the disk and the star is crucial to determining
the amount of stellar heating that occurs.

We assess the importance of the methods listed above for heating the
star in accreting pre-main-sequence stars and cataclysmic variables,
using recent models of boundary layers in these systems. We find that
intercepted radiation tends to be the most important source of stellar
heating in thin disk systems such as T Tauri stars and high-$\dot M$
cataclysmic variables. We argue that direct radiation across the
disk--star interface will be unimportant in steady-state systems.
However, it may be important in outbursting systems, where the
disk temperature rises and falls rapidly. Advection of thermal energy
into the star becomes the dominant source of stellar heating in thick
disk systems such as FU Orionis objects.

\end{abstract}


\section{Introduction}

Disk accretion onto a star is important in many contexts in
astrophysics. Viscous processes remove angular momentum from the disk
material, allowing it to spiral inward and eventually reach the
star. The gravitational potential energy released by the accreting
material is viscously dissipated, and most of it is radiated from the
disk surface.  Some of this radiation is intercepted by the accreting
star. Also, some energy and angular momentum remains in the accreting
material that reaches the surface of the accreting star.  We have
explored the problem of angular momentum accretion in disks elsewhere
(Popham \& Narayan 1991; Popham 1996; Popham \et 1996). Here we will
discuss the transfer of energy to the accreting star that results from
disk accretion.

Heating of the accreting star by disk accretion can have important
effects. In pre-main-sequence accretion disk systems such as FU
Orionis objects, heating due to disk accretion appears to expand the
stellar envelope. FU Orionis outbursts appear to arise in disks around
T Tauri stars, but accretion disk models for FU Orionis objects find
disk solutions where the stellar radii are 2--3 times larger than the
radii of T Tauri stars (Kenyon, Hartmann, \& Hewett 1988; Popham \et
1996). The expanded stellar radii result in significantly lower disk
luminosities and temperatures. The addition of accretion energy to
these stars will also alter their evolutionary tracks in the HR
diagram (Hartmann, Cassen, \& Kenyon 1996).

In cataclysmic variables, ultraviolet observations of some dwarf novae
have revealed that the white dwarfs are the dominant sources of
ultraviolet flux from these systems during quiescent periods.
Moreover, the white dwarfs in these systems have been observed to cool
and fade with time after an outburst ends (Hassall, Pringle, \&
Verbunt 1985; Verbunt \et 1987; Kiplinger, Sion, \& Szkody 1991; Long
\et 1994; G\"ansicke \& Beuermann 1996). The white dwarfs in these
systems are clearly heated during outbursts, and this implies that
disk accretion is responsible for the heating, since the accretion
rate and the disk temperature both increase dramatically during
outbursts.

Various authors have envisioned several ways in which an accretion
disk may heat the accreting star. The accretion disk has a luminosity
which in many cases greatly exceeds the luminosity of the accreting
star. Some fraction of the radiation from the disk surface will be
intercepted by the star (Adams \& Shu 1986).  Also, some fraction of
the energy dissipated in the innermost part of the disk may be
radiated inward across the disk--star boundary into the star (Bertout
\& Regev 1992; Lioure \& Le Contel 1994; Regev \& Bertout 1995).
Thermal energy may also be advected across this boundary by the
accreting material (Popham \& Narayan 1995; Popham \et 1996).

These methods for heating the star depend strongly on the structure of
the boundary layer between the disk and the star, where up to half of
the accretion luminosity will be released.  The proximity of the
boundary layer to the star means that a large fraction of the boundary
layer luminosity will be intercepted by the stellar surface.  The
temperature and density structure of the boundary layer region will
also determine whether the luminosity released there goes inward into
the star, outward into the disk, or simply upward to be radiated from
the disk surface. Finally, the efficiency of the boundary layer in
radiating away the energy dissipated there will determine the thermal
energy content of the accreting material flowing into the star.

Recently, we have constructed steady-state models of boundary layers
in cataclysmic variables (CVs) and accreting pre-main sequence stars
(Narayan \& Popham 1993; Popham \et 1993; Popham \& Narayan 1995;
Popham 1996; Popham \et 1996). These models follow the flow of the
accreting material through the disk and boundary layer and into the
outer layers of the accreting star. They are based on the slim disk
equations (Paczy\'nski \& Bisnovatyi-Kogan 1981; Muchotrzeb \&
Paczy\'nski 1982; Abramowicz \et 1988), which include a number of
terms which are ignored in the standard thin disk treatment. These
include radial pressure gradients and radial energy transport by
radiation and by advection of thermal energy. 

Our solutions directly calculate the radial distribution of flux from
the surface of the boundary layer and disk.  Most of the solutions
include a small settling zone inside the boundary layer which we take
to represent the outer layers of the star in an approximate way. The
radiative energy flux between this zone and the boundary layer gives
us some indication of the flux across the disk--star
interface. Finally, our solutions calculate the amount of thermal
energy advected into the star, based on the midplane properties of the
disk at the stellar surface. These features of our boundary layer
solutions permit us for the first time to calculate the importance of
each of these types of stellar heating by the accretion disk. We can
also compare stellar heating for different types of accreting stars.

Section 2 discusses heating of the star by intercepted disk
radiation. We calculate the disk luminosity intercepted by the star
for standard disk effective temperature profiles, and for our disk
models which include the boundary layer region.  In \S 3, we argue
that direct radiative heating across the disk--star boundary will be
unimportant in steadily accreting systems, but may be important in
outbursting systems where the boundary layer temperature changes on a
timescale comparable to the heating timescale. \S 4 discusses heating
by advection of thermal energy into the star. In \S 5, we compare the
relative importance of these types of stellar heating in various
accreting systems, including T Tauri and FU Orionis stars and
cataclysmic variables.

\section{Stellar Heating by Intercepted Disk Radiation}

\subsection{Stellar Heating for Standard Disk Temperature Profiles}

In most accretion disk systems, the accretion luminosity exceeds the
intrinsic luminosity of the accreting star, often by a large
factor. Therefore, if a substantial fraction of the radiation from the
disk is intercepted by the stellar surface, it can heat the star
to a significantly higher temperature than it would have in the
absence of disk accretion.

Adams \& Shu (1986) derived expressions for the heating of the star by
radiation from the disk. These are based on the assumptions that the
star is spherical and the disk is flat and infinitely thin. The
incident flux on the stellar surface is then
$$F_{inc}(\theta) = {\cos \theta \over \pi}{\int^{r_{out}}_{r_{min}}
\sigma T_d^4(r) r dr}{\int^{\phi_{max}}_{-\phi_{max}} (r \sin \theta
\cos \phi -1) (r^2 - 2 r \sin \theta \cos \phi + 1)^{-2} d \phi},$$
where $\theta$ is the polar angle of the point on the stellar surface,
$\phi$ is the azimuthal angle between this point and the point on the
disk, $T_d(r)$ is the effective temperature of the disk, $r$ is the
radius $R$ of the disk point in units of the stellar radius $R_*$,
$r_{out}$ is the radius of the outer edge of the disk, $r_{min} = 1 /
\sin \theta$ is the innermost radius of the disk seen by the point on
the star, and $\phi_{max} = \arccos [1/(r \sin \theta)]$ is the
largest azimuthal angle seen by the point on the star. For a given
variation of disk temperature with radius $T_d(r)$, we can integrate
to find the $\theta$-dependence of the incident flux on the stellar
surface and then integrate over the stellar surface to find the total
luminosity $L_{inc}$ intercepted by the star.

Adams \& Shu (1986) calculated that for a disk with a simple $T_d(r)
\propto r^{-3/4}$ temperature profile, $L_{inc} = (1/2 - 4/3\pi)
L_{disk} = 0.0756 L_{disk}$. Note that the standard thin disk
temperature distribution $T_d(r) \propto r^{-3/4} (1 -
r^{-1/2})^{1/4}$ reaches a maximum at $r \simeq 1.36$ and then
decreases in the innermost part of the disk (Shakura \& Sunyaev
1973). Numerically integrating the equation above with this
temperature distribution, we find that the luminosity intercepted by
the star is much smaller, only amounting to $L_{inc} = 0.0216
L_{disk}$. This difference clearly shows the importance of the inner
part of the disk in heating the star.  This suggests that when the
boundary layer luminosity is included in the effective temperature
distribution, a substantially larger fraction of the total accretion
luminosity will be intercepted by the star.

\subsection{Stellar Heating from Disk and Boundary Layer Radiation in
CVs and Pre-Main-Sequence Stars}

We can calculate the amount of stellar heating by disk radiation
directly, using numerical solutions for $T_d(r)$ in the boundary layer
and disk. Popham \& Narayan (1995) calculated boundary layer and disk
solutions for high-$\md$ cataclysmic variables. We select a typical
solution with $\md = 10^{-8} \msyr$, $\ms = 0.6 \msun$, shown in
Fig. 1 of Popham \& Narayan (1995). By numerically integrating the
equation given above, we find that the disk luminosity incident on the
stellar surface is $\linc \simeq 1.24 \times 10^{34} \ergs$, which is
a fraction 0.217 of the total disk luminosity of $5.73 \times 10^{34}
\ergs$. The reason for this very large incident luminosity is the
presence of the boundary layer, which radiates approximately half of
the total accretion luminosity from disk radii which are only a few
percent larger than $\rs$. Nearly half of this boundary layer
luminosity hits the stellar surface, so the overall fraction of the
total luminosity which reaches the stellar surface is close to $1/4$.

A similar situation occurs in T Tauri star accretion. Here we use the
$\md = 10^{-7} \msyr$ solution from Popham \et (1993) to calculate the
luminosity incident on the stellar surface, and find that $\linc
\simeq 1.39 \times 10^{33} \ergs$, which is 0.224 of the total disk
luminosity of $6.20 \times 10^{33} \ergs$. Again, this is due to the
boundary layer luminosity being radiated from a small region close to
the stellar surface.

In FU Orionis objects, the situation is quite different. Here, we use
a numerical solution from Popham \et (1996) which fits the spectral
energy distribution and line profiles observed from V1057 Cygni. This
solution has a high accretion rate $\md = 10^{-4} \msyr$, and as a
result the disk is quite thick, with the disk height $H \simeq 0.4
R$. The boundary layer luminosity is spread out over a large region of
the inner disk, so that a much smaller fraction of it reaches the
stellar surface. We find that $\linc \simeq 2.96 \times 10^{34}
\ergs$, which is only 0.0388 of the total disk luminosity of $7.62
\times 10^{35} \ergs$. In fact, this fraction of the disk luminosity
is only about half that which would reach the star from a disk with
$\teff\propto r^{-3/4}$, as discussed above. It is still larger than
the fraction from a standard thin disk, where no boundary layer is
included.

The reason for the large variation in the fraction of the disk
luminosity which reaches the stellar surface can be seen clearly in
Fig. 1a, where we show $T_d(r)$ for the solutions described above. The
CV and T Tauri solutions show an obvious peak in $T_d(r)$ near $R =
\rs$, whereas the FU Orionis solutions shows no such peak, since the
boundary layer luminosity is spread out over a large emitting area.

\subsection{The Fraction of the Disk Flux Incident on the Star}

For each disk radius, a certain fraction of the flux radiated from the
disk surface reaches the stellar surface. We can define a spherical
coordinate system centered at a point on the disk at radius $r$, where
the polar axis points toward the center of the star. We can then
integrate the flux from that point on the disk which is incident upon
the star. The star subtends a portion of the sky which extends out to
$\theta_0 = \arcsin(1/r)$, and the cosine of the angle to the normal to
the disk surface is $\sin \theta \cos \phi$. Thus we find the fraction
of the flux which hits the star to be
$$f_F(r) = {\int^{\theta_0}_0 \int^{\pi/2}_{-\pi/2} \sin \theta (\sin
\theta \cos \phi) d\phi d\theta \over \int^\pi_0 \int^{\pi/2}_{-\pi/2}
\sin \theta (\sin \theta \cos \phi) d\phi d\theta}
= {1 \over \pi}\left[\arcsin (1/r) - {(r^2 - 1)^{1/2} \over
r^2}\right].$$  
This is plotted in Figure 1b. 

We can integrate the flux which reaches the star $f_F F$ over the disk
surface to find the total luminosity incident on the star, and divide
this by the total disk luminosity to find the fraction of the disk
luminosity $f_L$ which reaches the star. This allows us to check the
values of $f_L$ which were calculated above by integrating the incident
flux over the stellar surface.  If we do this for a disk with $T_d(r)
\propto r^{-3/4}$, so that $F \propto r^{-3}$, we find
$$f_L = {\int^\infty_1 {1 \over \pi}\left[\arcsin (1/r) - {(r^2 -
1)^{1/2} \over r^2}\right] r^{-3} r dr \over \int^\infty_1 r^{-3} r
dr}$$
$$ = -{1 \over \pi} \left[{\arcsin(1/r) \over r} + {(r^2 -1)^{1/2}
\over r} + {(r^2 -1)^{3/2} \over 3 r^3}\right]^\infty_1 = {1 \over 2}
- {4 \over 3 \pi} = 0.0756.$$ 
This agrees with the value obtained above by integrating over the
stellar surface. We can follow the same procedure for the standard
thin disk flux distribution $F \propto r^{-3} (1 - r^{-1/2})$, and we
find $f_L = 0.0216$ as before.  Similarly, we can numerically
integrate the fluxes radiated from our disk and boundary layer
solutions for CVs and pre-main-sequence stars, weighting them by $f_F$
as given above, and again we find values of $f_L$ which agree with
those derived above.

\subsection{Distribution of Incident Flux on the Stellar Surface}

The different disk temperature distributions also produce rather
different distributions of the incident flux $F_{inc} (\theta)$ on the
stellar surface. $F_{inc}$ is always largest close to the equatorial
plane, and decreases to zero at the pole; however, the degree to which
$F_{inc}$ is concentrated toward the equator depends strongly on the
disk temperature distribution. In general, the incident flux
distribution on the star mirrors the flux distribution radiated by the
disk.  A disk which radiates a large flux from the inner disk or
boundary layer region will produce an incident flux distribution which
is concentrated near the equatorial plane. The incident flux
distributions produced by $T_d \propto r^{-3/4}$ and by standard thin
disk temperature distributions are shown in Figure 2a. They are
normalized by $F_* = \sigma T_*^4 = 3 G \ms \md / 8 \pi \rs^3$. Note
that $F_{inc}$ drops to zero at both $\theta = 0$ and $\theta =
\pi/2$. For the $r^{-3/4}$ distribution, $F_{inc}$ peaks only about
$0.6^\circ$ above the equatorial plane. For the standard thin disk
temperature distribution, which peaks at $r \simeq 1.36$, the
incident flux distribution peaks farther above the disk, at about
$7^\circ$ above the equatorial plane.

The distribution of incident flux on the stellar surface for our disk
and boundary layer solutions is shown in Figure 2b. The CV and T Tauri
disks have a sharp peak in the radiated flux near the stellar surface
due to the boundary layer, and these disks produce an incident flux
distribution on the stellar surface which peaks sharply near the disk
surface. The FU Orionis disk has a much smoother distribution of flux,
and the incident flux on the stellar surface is similarly rather
smooth. For comparison, we have plotted the incident flux on the
stellar surface which would result from disks with a standard
Shakura-Sunyaev temperature distribution, for the same values of
$\md$, $\ms$, and $\rs$. Over most of the surface of the star, the
incident flux distributions are nearly identical.  However, near the
equatorial plane, the boundary layer radiation increases the incident
flux dramatically, particularly in the CV and T Tauri cases. The
increase in the incident flux in the FU Orionis system is much
smaller, and is spread over a larger region of the stellar surface,
due to the larger radial extent of the boundary layer region.

The incident fluxes on the stellar surface shown in Fig. 2b are quite
large in some cases; to make them easier to compare to the intrinsic
fluxes from the accreting stars, we have included a vertical scale
showing the effective temperature corresponding to $F_{inc}$, $T_{inc}
= (F_{inc}/\sigma)^{1/4}$. For all three systems, $T_{inc}$ falls
somewhat below the peak boundary layer effective temperature; also
$T_{inc}$ exceeds the stellar temperature near the equatorial plane,
but drops below the stellar temperature away from the equatorial
plane.  For the CV solution, the peak value of $T_{inc}$ is $1.58
\times 10^5$ K. This is somewhat below the peak boundary layer
effective temperature of $2.25 \times 10^5$ K, but much hotter than
normal white dwarf temperatures, which are probably around $2 \times
10^4$ K. $T_{inc}$ stays higher than $2 \times 10^4$ K over most
of the stellar surface (Fig. 2b). For the T Tauri solution, the peak
$T_{inc}$ is about 6430 K, the peak boundary layer temperature is 8530
K and the stellar temperature is about 4000 K. Here $T_{inc}$ only
exceeds the stellar temperature in a small equatorial region, and
drops off rapidly away from the equator. For the FU Orionis solution,
the peak $T_{inc}$ is about 5790 K, the peak boundary layer
temperature is about 7110 K, and the stellar temperature can be
estimated at 5000 K. Again $T_{inc}$ only exceeds the stellar
temperature in a small equatorial zone, but here $T_{inc}$ falls off
more slowly with distance above the equatorial plane.

Note that in calculating the incident fluxes and luminosities, we have
continued to assume that the disk is flat and infinitesimally
thin. This is a fairly good approximation for the CV and T Tauri
disks. These are quite thin, and while they have a concave surface,
this should only produce substantial changes in the flux coming from
the outer parts of the disk, which make a very small contribution to
the total flux reaching the star. The FU Orionis disk has a fractional
thickness $H/R \sim 0.4$, where $H$ is the vertical pressure scale
height; the photosphere of the disk is probably a few scale heights
above the midplane. Thus, in this case the disk phtotsphere may have a
conical shape, and the disk will intercept radiation from
itself. Since the disk will presumably cover much of the star, only
a small portion of the stellar surface will intercept radiation from
the disk.

\section{Direct Radiative Heating Across the Disk--Star Boundary}

\subsection{The Inner Boundary Condition}

Direct radiative heating of the star across the disk--star boundary is
the most problematic of the heating terms we consider here. Current
boundary layer models assume a steady state, making it difficult to
calculate the radial energy transport between the boundary layer and
the star.  The steady-state assumption requires that the mass
accretion rate $\dot M$ be constant for all radii.  This is a
reasonable assumption to make for the disk and boundary layer, but it
is probably not realistic for the accreting star, since the accreted
matter may accumulate at the surface of the star or cause changes in
the stellar radius.  Since steady-state boundary layer models cannot
treat the star realistically, in general the inner edge of the
computational grid is placed at or near the boundary layer--star
interface. The radiative energy transport across the inner boundary
must then be specified by a boundary condition, but it is not obvious
what condition to use.

In our models (Narayan \& Popham 1993; Popham \et 1993; Popham \&
Narayan 1995), we try to avoid this problem by setting the inner
boundary of our calculation just below the stellar surface, at $R
\simeq 0.95 - 0.98 R_*$. The region between the inner boundary and the
BL--star interface at $R=R_*$ is a pressure-supported settling flow
rotating at the stellar rotation rate, with the same $\dot M$ as the
disk and boundary layer; we take this settling region to represent the
outer layers of the accreting star. We impose a boundary condition on
the radially-directed radiative energy flux at the inner boundary,
which matches the flux from the surface of the star in the absence of
accretion. However, the presence of the settling region between the
inner boundary and the boundary layer isolates the boundary layer from
the inner boundary condition on the radial flux. Thus, the radiative
energy transport across the BL--star interface is determined primarily
by the conditions in the settling region, unless the radial luminosity
entering or leaving at the inner boundary is comparable to the
boundary layer luminosity. In general, our solutions have flux
radiated from the settling region into the boundary layer, but the
luminosity across the BL--star interface is generally much less than
the accretion luminosity, so that the heating of the boundary layer by
the star is insignificant. The BL luminosity in these solutions
travels radially outward into the inner disk as it diffuses toward the
disk surface, producing a ``hot'' boundary layer. Thus, in our
steady-state models, there is no heating of the star by radiative flux
across the boundary layer--star interface.

Other steady-state boundary layer models (Bertout \& Regev 1992;
Lioure \& Le Contel 1994; Regev \& Bertout 1995) have put a boundary
condition on the temperature of the accreting material at the boundary
layer--star interface at $R=\rs$, which they take to be equal to the
assumed photospheric temperature of the star in the absence of
accretion. In general, this is substantially lower than the
temperature of the boundary layer gas, so that there is a strong
temperature gradient with the temperature decreasing inward. This
produces a large inward flux of radiative energy, which can carry a
large fraction of the energy dissipated in the boundary layer into the
star. Solutions for boundary layers around T Tauri stars which use
this boundary condition have found that the luminosity radiated into
the star $L_{in}$ is a substantial fraction of the total accretion
luminosity $L_{acc}$; Bertout \& Regev (1992) found $L_{in} = 0.2
L_{acc}$, Lioure \& Le Contel (1994) $L_{in} = 0.286 L_{acc}$, and
Regev \& Bertout (1995) $L_{in} = 0.3 L_{acc}$. These luminosities
represent an even larger fraction of the rate at which energy is
dissipated in the boundary layer. For an accreting star rotating at a
fraction $f$ of the Keplerian velocity, the energy dissipation rate in
the boundary layer is $\simeq 0.5(1-f)^2 L_{acc}$. All of the solutions
discussed use $f=0.1$, so that the dissipation rate is $\simeq 0.4
L_{acc}$, and the fraction of the dissipated energy which is radiated
into the star ranges from $\simeq 0.5 - 0.75$. As a result, these are
``cool'' boundary layer solutions, where the boundary layer effective
temperature is substantially lower than it would be in solutions where
little or no luminosity goes into the star.

Godon, Regev, \& Shaviv (1995) suggested that the differences between
the two types of solutions are due to a fundamental difference between
placing boundary conditions on the flux and placing them on the
temperature. They compared the effects of flux and temperature
boundary conditions using four boundary layer solutions for accretion
disks in CVs. Two of the solutions used an inner boundary condition on
the midplane temperature at $\rs$, $T_* = 2 \times 10^5$ K, and the
other two instead used a condition on the radial radiation flux at
$\rs$, $F_* = \sigma \teff^4$, where $\teff = 2 \times 10^4$ K. The
angular velocity profiles of the solutions show little dependence on
the type of boundary condition used. The temperature profiles differ;
in all of the solutions, the temperature of the accreting material
rises as it moves inward, and exceeds $2 \times 10^5$ K. In the
solutions with a temperature boundary condition, the temperature then
drops back down to the assumed boundary value of $2 \times 10^5$ K at
$\rs$, while in the solutions with a flux condition, the temperature
continues to rise gradually inward to $\rs$. Godon, Regev, \& Shaviv
(1995) claim that this difference reflects the existence of two
different families of solutions.

However, it seems clear that if appropriately higher values of $T_*$
were chosen, the resulting solutions would be identical to the
flux-condition solutions, and conversely, if an appropriate negative
radial flux (carrying energy into the star) were chosen, the
flux-condition solutions could be made identical to the
temperature-condition solutions shown by Godon, Regev, \& Shaviv
(1995). We have verified this by calculating two solutions, one using
a flux condition, and the other a temperature condition where the
temperature has been chosen to match that produced by the flux
solution. The resulting solutions are identical in all respects. Thus
the nature of the solutions depends not on the type of boundary
condition, but rather whether the flux or temperature they impose
results in an inward or an outward radiative flux.

Solutions which have a large inward flux, whether it results from a
boundary condition on the flux or the temperature, face a serious
problem. The inward flux results from the temperature decreasing
inward. This flux cools the boundary layer by transferring heat to the
star, but in doing so, it rapidly heats the outer layers of the
star. This reverses the temperature gradient that produced the inward
flux in the first place. For instance, if the boundary layer--star
interface is assigned the photospheric temperature in the absence of
accretion, it is cooler than both the boundary layer and the stellar
envelope.  Thus, it represents a minimum in the temperature as a
function of radial distance from the center of the star.  Therefore,
flux from both the star and the boundary layer will heat this region;
radiative diffusion will fill in the minimum in the temperature. If a
large fraction of the boundary layer luminosity goes into the star,
the outer layers of the star will be heated until the temperature
gradient is reversed. Here we show that in fact, this heating will
occur quickly, so that it is unrealistic to use a boundary condition
that results in a large inward flux in steady-state boundary layer
models.

\subsection{The Heating Timescale}

One can estimate the time to heat the outer layers of the star as
follows. The temperature gradient, and therefore the flux, will go to
zero when the outer layers of the star are heated to the same
temperature as the boundary layer gas, $T_{BL}$. Thus, if we take the
mass $m_{heat}$ of the portion of the star which is cooler than
$T_{BL}$, we can derive the energy required to heat that mass of gas
to $T_{BL}$. From this, we can estimate the heating timescale; in
general, the boundary layer luminosity should be the dominant source
of heating. Assuming that all of the boundary layer luminosity goes
into heating the boundary layer--star interface region, we find a
heating timescale 
$$t_{heat} = c_P T_{BL} m_{heat} / L_{BL}.$$

First, we must estimate $m_{heat}$. This can be done simply by taking
the standard equation for diffusion of radiative energy, and
transforming it so that the mass $m$ is used as the independent
variable. This gives
$${dT \over dm} = -{3 \kappa T_{eff}^4 \over 64 \pi R^2 T^3},$$ 
where $T_{eff}$ is the effective temperature of the star. This
gives us an estimate for $m_{heat}$,
$$m_{heat} \simeq \Delta m \simeq {64 \pi R^2 T^3 \over 3 \kappa
T_{eff}^4} \Delta T \simeq {16 \pi R^2 \over 3 \kappa T_{eff}^4}
\Delta (T^4) \simeq {16 \pi R_*^2 T_{BL}^4 \over 3 \kappa
T_{eff}^4}.$$ 

This estimate for the mass to be heated assumes that the entire
surface of the star will be heated by the boundary layer. In fact only
the equatorial region will be heated directly. The remainder of the
stellar surface would have to be heated indirectly, by radiative
diffusion or fluid motions which would transfer heat from the
equatorial region toward the poles. Radiative diffusion is unlikely to
be able to accomplish this, since the radiative energy will escape
from the stellar surface as it moves toward the poles. Since the
boundary layer heating of the equatorial region only extends to a
modest depth in the stellar atmosphere, the timescale for diffusion to
the surface is probably much shorter than that for diffusion to the
poles. Diffusion of the radiative energy by fluid motions would suffer
from the same problem; in order to spread the boundary layer heating
evenly around the stellar surface, these motions would have to move
the heated material to the poles before it could cool. Here we
consider both alternatives: heating of the surface layers of the
entire star, or of only the equatorial regions. The boundary layer
extends to a height $H_{in}$ above and below the equatorial plane, so
it directly heats an area $4 \pi R_* H_{in}$ of the stellar surface,
which is a fraction $H_{in} / R_*$ of the total stellar surface
area. Therefore, for heating of the equatorial regions only, we reduce
the estimate of the mass to be heated by this factor.

\subsection{Heating Timescales for CVs and Pre-Main-Sequence Stars}

We can use the boundary layer temperatures taken from our numerical
solutions to estimate heating timescales for CVs and accreting
pre-main-sequence stars.  For a cataclysmic variable, we take as an
example the solution with $\md = 10^{-8} \msyr$, $\ms = 0.6 \msun$,
$\oms=0$, $\alpha = 0.1$, presented in Popham \& Narayan (1995). The
white dwarf has a radius $\rs = 8.7 \times 10^8 {\rm ~cm}$, and we
adopt a white dwarf temperature $T_{eff} = 10,000$ K. The boundary
layer has $T_{BL} \simeq 350,000$ K and $H_{in} / R_* \simeq
0.025$. If we again assume $\kappa = 1 \cm2g$, we find $m_{heat}
\simeq 2 \times 10^{25} \g$ if the whole stellar surface is heated. To
heat this amount of gas to $350,000$ K will require $\sim 2.4 \times
10^{39}$ ergs. The accretion luminosity for these parameters is $\sim
6 \times 10^{35} \ergs$, so half of the accretion luminosity will heat
the gas in $8 \times 10^4 {\rm ~s} \simeq 1$ day. If only the
equatorial region is heated, we find $m_{heat} \simeq 5 \times 10^{23}
\g$, and a heating timescale of $2 \times 10^3 \s$.

For a T Tauri star, we adopt typical values of $R_* = 2 ~R_\odot$,
$T_{eff} = 4000$ K. We take $T_{BL} \simeq 20,000$ K and $H_{in} / R_*
\simeq 0.06$ from a T Tauri boundary layer solution with $\dot M =
10^{-7} \msyr$, $M_* = 1 \msun$, $\oms = 0$, $\alpha=0.1$ (Popham \et
1993). The opacity $\kappa$ varies strongly with temperature and
density in this regime. The boundary layer solution gives densities
$\rho \simeq 10^{-7}-10^{-8} \gcm3 $ in the inner boundary layer. At
$T = 20,000$ K, this would give $\kappa \sim 100 \cm2g$ or more, but
at lower temperatures, the opacity drops off rapidly, reaching $\kappa
\sim 0.01 \cm2g$ at $T = 4000$ K (Alexander \& Ferguson 1994). If we
assume for simplicity that $\kappa = 1 \cm2g$, we find $m_{heat}
\simeq 2 \times 10^{26} {\rm ~g}$ if the whole stellar surface is
heated (if we instead approximate the opacity as a power law $\kappa =
10^{-38} T^{10}$, we find $m_{heat} \simeq 2 \times 10^{25} {\rm
~g}$). To heat this amount of gas to $T_{BL} = 20,000$ K requires
energy $c_P T_{BL} m_{heat} \simeq 1.4 \times 10^{39} {\rm
~ergs}$. The accretion luminosity for these parameters is $L_{acc} = G
M_* \dot M / R_* \simeq 5.6 \times 10^{33} \ergs$. If we assume that
half of this luminosity goes into heating the star, we find a heating
timescale $t_{heat} \simeq 5 \times 10^5 {\rm ~s} \simeq 6$ days. If
only the equatorial regions are heated, we find $m_{heat} \simeq 1.2
\times 10^{25} \g$, and a heating timescale of $3 \times 10^4 \s
\simeq 0.35$ days.

In FU Orionis objects, the heating timescale is much longer than in T
Tauri stars, due primarily to the much higher boundary layer
temperature. Our solution for V1057 Cygni has a midplane temperature
of $T_{BL} \simeq 2 \times 10^5$ K. Therefore the mass of the portion
of the stellar envelope which has a temperature below $T_{BL}$ is much
larger; taking $\rs = 3.5 \times 10^{11}$ cm and the star's effective
temperature to be $T_{eff} = 5000$ K, we find $m_{heat} \simeq 5
\times 10^{30}$ g. Half of the accretion luminosity is $0.5 L_{acc}
\simeq 6.3 \times 10^{35} \ergs$, so the heating timescale is
$t_{heat} \simeq 5.7 \times 10^8 \sec \simeq 18$ years.

The estimates derived here are quite rough. They depend on the fourth
power of the ratio $T_{BL}/T_{eff}$, so they are quite sensitive to
variations in these temperatures. The boundary layer temperature can
vary significantly for different values of $\md$, $\ms$, $\oms$, and
$\alpha$ (Popham \& Narayan 1995).  They also assume that the dominant
means of energy transfer in the boundary layer--star interface region
is radiative diffusion. In some cases (e.g. T Tauri stars), the
accreting star may have a convective envelope.  If convection is
present in the heated zone, it would lead to a different expression
for the heating timescale, and it could produce more efficient
sideways transport of the boundary layer energy within the
star. However, the assumed boundary layer heating will reverse the
direction of energy transfer in the adjoining portion of the stellar
envelope, which should suppress convection in this region. Prialnik \&
Livio (1985) found that for simple spherical accretion at moderate
rates onto an initially fully convective 0.2 $M_\odot$ star, the
convective zone recedes from the stellar surface. For accretion rates
of $10^{-7} \msyr$ or below, their results indicate that this occurs
roughly when the inward energy flux due to accretion exceeds the
outward flux from the star's inherent luminosity. A T Tauri star has
an effective temperature around 4000 K, so the stellar flux is about
$1.5 \times 10^{10} \ecs$. If the boundary layer luminosity, $0.5
L_{acc} = G M \md / 2 \rs \simeq 3 \times 10^{33} \ergs$ is radiated
into the equatorial region, which has a surface area $4 \pi \rs H_{in}
\simeq 1.5 \times 10^{22} {\rm ~cm}^2$, the inward flux of accretion
energy is $\simeq 2 \times 10^{11} \ecs$. This is much greater than
the stellar flux, and should be sufficient to suppress convection in
the equatorial surface layers of the star.

\subsection{Direct Radiative Heating in Steadily Accreting Systems}

The heating timescales derived above for CV and T Tauri accretion
disks are very short. If the mass accretion rate remains steady over
this short timescale, the stellar envelope will be heated to the
boundary layer temperature, and the inward radiative flux will vanish
as the temperature gradient flattens out. The system should then reach
a steady state where there is no inward radiative flux across the
disk--star boundary.

The use of a boundary condition which sets the temperature at the
disk--star interface to the photospheric temperature of the
unperturbed star inevitably results in a large inward flux, since the
temperature is forced to drop to an unnaturally low value, and the
resulting temperature gradient forces the radiative flux to be
directed inward (Lioure \& Le Contel 1994; Regev \& Bertout 1995).
The short heating timescales for the stellar envelope show that this
temperature gradient would quickly be eliminated by the rapid heating.
Thus, such boundary conditions (or, equivalently, boundary conditions
which directly specify a large inward flux) should not be applied to
steadily accreting systems.

The use of a flux boundary condition offers a significant advantage
over a temperature condition. If the radial flux is chosen to have a
small value and be directed outward, the resulting temperature
gradient will be small, and there will be no minimum in the
temperature at the disk--star boundary. If a temperature condition is
used, one must guess the value of the temperature at the disk--star
boundary, and even a relatively small error in the temperature can
result in a large radial flux, which can heat or cool the boundary
layer and change its structure substantially.

\subsection{Reradiation of the Boundary Layer Luminosity by the
Stellar Surface Layers}

One way around the argument given above is to say that the surface
layers can radiate the boundary layer luminosity away at the same rate
as it enters, so that a steady state will be reached, as argued by
Regev \& Bertout (1995). However, for the boundary layer luminosity to
be radiated inward toward the star and then reradiated from the
stellar surface would require that the optical depth along this
indirect path be lower than the vertical optical depth of the boundary
layer itself. This is not the case, for the simple reason that below
the boundary layer, the accreting material must be supported by the
pressure gradient, so that the pressure and density of the accreting
gas increase rapidly with decreasing radius. By contrast, the density
is low immediately outside the boundary layer due to the large radial
velocity of the accreting material. As a result, the vertical optical
depth of the accreting material is smallest just outside the boundary
layer, and increases rapidly as the material flows inward and settles
onto the star. This leads to the natural conclusion that the boundary
layer luminosity is radiated radially outward rather than inward, and
emitted directly from the surface of the boundary layer and inner disk
(the ``thermal boundary layer''), rather than indirectly from the
surface of the star. This is exactly what happens in our boundary
layer solutions.

Regev \& Bertout (1995) argued that most of the boundary layer
luminosity in T Tauri stars would be radiated into the star and
subsequently radiated vertically, away from the equatorial plane,
finally emerging from the stellar surface above and below the
disk. They first calculated the temperature distribution of the
stellar atmosphere in the equatorial plane, using an expression for
the variation of temperature in an atmosphere illuminated from
above. However, this expression is inappropriate for calculating the
temperature in the equatorial plane, since it treats the boundary
layer only as a source of incident radiation, and assumes that the
equatorial surface layers of the star can radiate freely into
space. In fact, the equatorial surface layers of the star are buried
beneath the boundary layer and disk, so the optical depth in the
equatorial plane is very large. Regev \& Bertout also claimed that the
indirect emission from the stellar surface layers would be
observationally indistinguishable from direct emission from the
boundary layer. This was based on assuming that the vertical optical
depth of the emitting region would be about one, but boundary layer
solutions show that the vertical optical depth increases rapidly as
the material moves inward through the boundary layer, and is much
larger than one.

\subsection{Direct Radiative Heating in Outbursting Systems}

Although direct radiative heating of the star by the boundary layer is
unlikely to be important in steady accretion, it could be significant
in systems where the accretion rate varies on a short timescale.  A
number of accreting systems, such as dwarf novae and FU Orionis
objects, experience outbursts where their brightness increases quite
rapidly. The outbursts in these systems are generally interpreted as
being due to an increase in the mass accretion rate, which would in
turn increase the boundary layer temperature. If the outburst rise
time is faster than the heating timescale of the stellar envelope
derived above, then the hot boundary layer will radiatively heat the
star.

For dwarf novae, the rise times of outbursts are generally around 1
day, comparable to the heating timescales derived above. This suggests
that during the early phases of the outbursts, a large fraction of the
boundary layer luminosity could go into heating the white dwarf
surface layers. This could have two important effects. First, the
increase in the boundary layer luminosity would be delayed until the
white dwarf surface layers were heated, which might account for some
of the observed delay between the optical and EUV fluxes during
outbursts (e.g., Mauche, Raymond, \& Mattei 1995). Second, the energy
stored in the heated surface layers of the white dwarf would be
radiated away during quiescence, as observed in a number of cases
(Hassall, Pringle, \& Verbunt 1985; Verbunt \et 1987; Kiplinger, Sion,
\& Szkody 1991; Long \et 1994; G\"ansicke \& Beuermann 1996). This is
discussed further in \S 5.2.

For FU Orionis objects, we derived a heating timescale of $\sim 18$
years in \S 3.3.  This is substantially longer than the rise
times of FU Orionis and V1057 Cygni, which are $\sim 1$ year, and is
probably a substantial fraction of the length of the outbursts, which
are estimated to last $\sim 100$ years. Thus we expect that
direct radiative heating across the star--disk boundary will heat the
stellar envelope in the early part of the outburst, until the envelope
temperature reaches $T_{BL}$. However, the fact that the stellar radii
derived from disk models of FU Orionis objects are substantially
larger than those of T Tauri stars suggests that the star expands
rapidly in response to the high outburst accretion rate. Thus, the
interaction between the disk and star is probably more complicated
than our simple picture of radiative heating. In particular, the
``stellar envelope'' may in fact consist largely of accreted material
which is already at a high temperature, so that little or no radiative
heating would take place across the disk--star boundary.

\section{Heating by Advection of Thermal Energy}

The material accreted by the star carries thermal energy with it. In
some cases this can be an important source of stellar
heating. Advection of entropy is included in the slim disk energy
equation (Paczy\'nski \& Bisnovatyi-Kogan 1981; Muchotrzeb \&
Paczy\'nski 1982; Abramowicz \et 1988). It may have an important
effect in accretion disks around black holes, since thermal energy
which is advected into the black hole will not be radiated. This
changes the luminosity and spectral energy distribution of the black
hole system (Narayan, McClintock \& Yi 1996). In accretion onto a
star, the thermal energy of the accreting material cannot be
eliminated in this way. Instead, it is advected into the star.

A good indicator of the importance of energy advection in the disk is
the vertical pressure scale height $H$ of the disk. The standard
expression for the disk height is $H \simeq c_s / \ok$, which can be
rewritten $H^2 / R^2 \simeq c_s^2 / v_K^2$. Thus the square of the
disk height as a fraction of the disk radius $R$ is proportional to the
thermal energy content of the accreting material, as a fraction of the
gravitational energy released $G \ms \md / R = \md v_K^2$. Therefore,
we should expect that energy advection will be important in thick
disks, like those found in systems with high accretion rates.

\subsection{The Fraction of the Accretion Luminosity Advected Into the
Star}

A more accurate calculation of the energy advected into the star can
be made by using the slim disk energy equation, one of the equations
of our boundary layer model, which reads
$$
\nu \Sigma \left ( R{d\Omega \over dR}\right ) ^2 - F_V -\Sigma v_R
T_c {dS \over  dR} - {1 \over R}{d \over dR}(RHF_R) = 0.
$$
The first term in this equation represents the viscous dissipation
within the disk; $\nu$ is the kinematic viscosity, $\Sigma$ is the
surface density of the disk, and $d \Omega / dR$ is the angular
velocity gradient. The second term is the vertical flux $F_V$ from the
disk surface. The third term is the advected entropy; $v_R$, $T_c$,
and $S$ are the radial velocity, midplane temperature, and entropy of
the accreting material. The final term is the divergence of the radial
radiation flux $F_R$ between adjacent disk annuli. If we
integrate this equation over the surface of the disk, we have
$$
L_{diss} - L + \md {\int T_c dS} - 4 \pi [R_{out} H_{out} F_{R,out} -
\rs \hs F_{R,*}] = 0.
$$
Thus the overall energy balance is determined not only by the rate of
energy dissipation within the disk $L_{diss}$ and the rate at which
energy is radiated away from the disk surface $L$, but also by the
energy carried into and out of the disk by energy advection and radial
radiation.  

We can use the other slim disk equations, in particular the angular
momentum equation and the radial momentum equation, to derive an
expression for $L_{diss}$ (see Popham \& Narayan 1995 for details),
$$
L_{diss} = {G \ms \dot M \over \rs} \left( 1 - j {\oms \over \omks} +
{1 \over 2} {\oms^2 \over \omk2s} \right) 
+ \dot M {\int {dP \over \rho}}.
$$
The first term is the accretion luminosity $L_{acc} = G \ms \md /
\rs$, corrected for the rotational kinetic energy transferred between
the star and the disk, where $j \equiv \dot J / \dot M \omks \rs^2$ is
the angular momentum accretion rate in standard units, and $\oms$ and
$\omks$ are the stellar rotation rate and the Keplerian rotation rate
at $R_*$. The second term accounts for radial pressure support of the
accreting material; $P$ is the pressure.

The radial radiation fluxes at the inner and outer edges of the disk
are small. At the inner edge, as discussed above, we assume that the
flux entering the disk is $\sigma T_*^4$, where we take $T_*$ to be
the effective temperature of the unperturbed star. Thus a fraction
$\hs / \rs$ of the star's luminosity goes into the disk, where $\hs$
is the disk height at the stellar surface. In general, the star's
luminosity is much smaller than $L_{acc}$, and in some cases $\hs \ll
\rs$, so the luminosity entering the inner edge of the disk is
insignificant, and at the outer edge the radial flux is
negligible; thus we can neglect these radial radiation terms. 

We can then use $T dS = dU - P d \rho /
\rho^2$ and the expression for $L_{diss}$ given above to write an
expression for the luminosity radiated by the disk
$$
L = L_{acc} \left( 1 - j {\oms \over \omks} + {1 \over 2} {\oms^2
\over \omk2s} \right) + \md {\int {dP \over \rho}} + \md {\int dU} -
\md {\int {P \over \rho^2} d\rho}
$$
$$
= L_{acc} \left( 1 - j {\oms \over \omks} + {1 \over 2} {\oms^2
\over \omk2s} \right) + {5 \over 2} \md {\int d \left({P \over
\rho}\right)},
$$
$$
= L_{acc} \left( 1 - j {\oms \over \omks} + {1 \over 2} {\oms^2
\over \omk2s} \right) - \md c_P T_c(\rs)
$$
where the internal energy $U = 3 P / 2 \rho$, and $T_c(\rs) \gg
T_c(\ro)$. The first term is the total energy dissipated in the disk;
$\md c_P T_c(\rs)$ of this is advected into the star, and the
remainder is radiated away. 

It is easy to see how the square of the disk thickness at $\rs$
is a direct measure of the importance of advection. The rate of energy
advection into the star is $\md c_P T_c(\rs) \simeq 2.5 \md
c_s^2(\rs)$, while the accretion luminosity is $G \ms \md \rs = \md
\ok^2(\rs)$. Thus a fraction $f = 2.5 c_s^2(\rs)/\ok^2(\rs) = 2.5
(\hrs)^2$ of the accretion luminosity is advected into the star.

\subsection{Advective Heating in CVs and Pre-Main-Sequence Stars}

Using boundary layer solutions, we can find the midplane temperature
at $\rs$ for various types of accretion disks. For a high-$\md$ CV
solution taken from Popham \& Narayan (1995) (with $\md = 10^{-8}
\msyr$, $\ms = 0.6 \msun$, as discussed above), we find $T_c(\rs)
\simeq 7 \times 10^5$ K, and $\hrs \simeq 3.3 \times
10^{-2}$. Thus thermal energy is advected into the white dwarf at the
rate of $1.5 \times 10^{32} \ergs$, or about $2.6 \times 10^{-3}
\lacc$.
 
To calculate the energy advection into T Tauri stars, we use a
boundary layer solution with $\md = 10^{-7} \msyr$ from
Popham \et (1993). This solution has $T_c(\rs) \simeq 2 \times 10^4$
K, and $\hrs \simeq 5.7 \times 10^{-2}$. Therefore the rate of
energy advection into the star is about $4 \times 10^{31} \ergs$,
which is $\sim 8 \times 10^{-3} \lacc$.

In FU Orionis objects, the situation is quite different. Here the disk
is quite thick as a result of the high accretion rates $\md \simeq
10^{-4} \msyr$ reached during FU Orionis outbursts.  The optical depth
from the disk surface to the midplane is very large, so the midplane
temperatures are high. In the solutions for FU Orionis and V1057 Cygni
calculated by Popham \et (1996), $T_c(\rs) \simeq 2 - 2.5 \times 10^5$
K, and $\hrs \simeq 0.39$, about 10 times as large as in the CV and T
Tauri solutions.  Since the fraction of the accretion luminosity
advected into the star goes as the square of $\hrs$, it is on the
order of 100 times larger than in the thin disks. For $\hrs = 0.39$,
we have $2.5 (\hrs)^2 \simeq 0.38$, so 38\% of the accretion
luminosity is carried into the star. The accretion luminosities of
these systems are quite large, generally around $1 - 3 \times 10^{36}
\ergs$, so the rates of energy advection are also very large, around
$4 - 12 \times 10^{35} \ergs \simeq 100 - 300 \lsun$. Of course, these
rates are only reached during FU Orionis outbursts.

\section{Discussion}

\subsection{The Relative Importance of Various Sources of Stellar
Heating}

In the preceding sections, we have calculated the rates of various
types of stellar heating in different accreting systems. The types of
stellar heating we have considered include heating by radiation
emitted from the disk surface, direct radiative heating through the
disk--star boundary, and advection of thermal energy. The accreting
systems for which we analyze the stellar heating rates include
steadily accreting thin disk systems such as cataclysmic variables and
T Tauri stars, thick disk FU Orionis systems, and systems undergoing
outbursts such as dwarf novae and FU Orionis systems.

We find that in steady thin disk systems, radiation emitted from the
disk surface is the most important source of stellar heating. In CVs
and T Tauri stars, we find that over 20\% of the accretion luminosity
reaches the star. Most of this comes from the boundary layer region.
Advection of thermal energy is unimportant in these systems, since
almost all of the accretion energy is radiated away, leaving very
little thermal energy in the accreting material.

In steady thick disk systems, which may apply to some FU Orionis
objects, the situation is quite different. The dominant source of
stellar heating is advection of thermal energy into the star. A large
fraction of the accretion luminosity is stored in the accreting
material, resulting in a thick disk, and carried into the
star. Heating by radiation from the disk surface is much less
important than in the thin disk systems. Since the boundary layer
luminosity is emitted over a wide range of radii, a much smaller
fraction of it reaches the stellar surface.

We have argued that direct radiative heating through the disk--star
boundary will be insignificant in steadily accreting systems, since
any temperature gradient which would cause the boundary layer
luminosity to flow inward will be reversed on a short timescale. In
outbursting systems, direct radiative heating can become important if
the timescale on which the boundary layer temperature rises is shorter
than the heating timescale of the stellar envelope. Direct heating of
the star will continue until the envelope is heated to the new
boundary layer temperature or the outburst ends. Thus we expect that
direct heating may be important during the early portion of both dwarf
nova and FU Orionis outbursts.

In this paper, we have concentrated on external sources of stellar
heating, which result from the transfer of radiation or hot material
from an accretion disk onto the surface of the star. Additional
heating may result from processes which occur within the star as a
result of accretion, such as compressional and shear heating.
Compressional heating of the star results from the change in the
stellar structure due to the mass added by accretion. It should be
particularly important in white dwarfs, which become much more
centrally condensed as their masses increase. Sion (1995) has computed
the effects of this heating in white dwarfs during dwarf nova
outbursts with accretion rates of $10^{-8} - 10^{-7} \msyr$. He found
that compressional heating alone can produce variations of up to 5000
degrees K in the effective temperature of the white dwarf from
outburst to quiescence. Shear heating will take place as the internal
rotation of the star adjusts to the torque applied to the surface
layers by accretion. Since both of these processes take place within
the interior of the star, they depend less strongly on the details of
the disk structure, and take place over a longer timescale than the
external heating sources we have discussed in this paper.

\subsection{Effects of Heating on the Star}

The main focus of this paper is on determining the importance of
various types of stellar heating in different types of accreting
systems. Nonetheless, it seems appropriate to briefly consider the
effects of this heating upon the accreting star.  All of the types of
stellar heating described in \S 2--4 above heat the star from outside,
but do so in rather different ways. 

The external illumination of the stellar surface by radiation from the
disk, described in \S 2, will heat the atmosphere of the star.  The
effects of external illumination of a stellar atmosphere have been
studied by a number of authors, many of whom have focused on
irradiation of the donor star in accreting X-ray binary systems. If
the incident radiation is in the form of X-rays, the heating due to
external illumination may drive a wind from the surface of the star
(see, e.g., Tavani \& London 1993 and references therein).  In CVs,
much of the boundary layer emission will be in X-rays, and as we
discussed in \S 2, a significant fraction of the boundary layer
emission will be intercepted by the equatorial regions of the star.
Tavani \& London (1993) found that incident X-ray fluxes $\simeq
10^{15} \ecs$ on a white dwarf drove winds with mass loss rates $\sim
10^{18} \g \pers$. As Fig. 2b demonstrates, incident X-ray fluxes of
this magnitude are produced by the boundary layer radiation in CVs,
but only in the equatorial region of the accreting white dwarf. Since
Tavani \& London's (1993) mass loss rates were based on spherically
symmetric winds, and because those winds arose from an extremely
low-mass, large-radius ($M_{WD} = 0.1 \msun$, $R_{WD} = 5 \times 10^9
{\rm cm} $) white dwarf, the mass loss rates from CVs should be
expected to be quite a bit smaller. Nonetheless, irradiation of the
equatorial regions of the white dwarf could make an important
contribution to driving winds from CVs.

In pre-main-sequence stars, the external illumination will be in the
form of much softer optical and ultraviolet radiation. Most studies of
irradiated atmospheres in these systems have focused on the
illumination of the disk atmosphere by radiation from the star (Calvet
\et 1991; Malbet \& Bertout 1991), the opposite situation to the one
described here. However, these authors treated the disk atmosphere in
a way that should be reasonably applicable to the irradiation of the
star by the disk. These studies found that irradiation heats the outer
layers of the atmosphere, producing a sort of ``chromosphere''. This
heated upper layer can produce emission lines which can fill in the
absorption lines of the unperturbed stellar atmosphere.

Direct radiative heating across the disk--star interface during disk
outbursts will heat an equatorial region which sits at a larger
optical depth from the surface. When the outbursts ends and the disk
temperature drops, this heated equatorial region will cool. A decline
in the ultraviolet flux during quiescence has been observed in several
dwarf nova systems (Hassall, Pringle, \& Verbunt 1985; Verbunt \et
1987; Kiplinger, Sion, \& Szkody 1991) and has been confirmed to be
due to cooling of the white dwarf (Long \et 1994; G\"ansicke \&
Beuermann 1996).

Pringle (1988) modeled the heating and cooling of the white dwarf
surface layers which should occur during a dwarf nova outburst and
subsequent quiescent phase. He used a range of temperatures for the
external heat source, and varied this external temperature with time
to simulate a dwarf nova outburst. Pringle found that the assumed
temperature of the heat source had a dramatic effect on the timescale
for cooling, since higher temperatures heated the white dwarf to a
greater depth. A fairly high external temperature $\sim 900,000$ K was
required to make the cooling last long enough to agree with
observations by Verbunt \et (1987) of the decrease in the UV flux of
VW Hydri during quiescence.

In the discussion above, we cited the example of a $0.6 \msun$ white
dwarf accreting at $10^{-8} \msyr$, for which our boundary layer
solution gave $T_{BL} \simeq 350,000$ K.  This temperature is
substantially lower than the temperature of $900,000$ K cited by
Pringle (1988), and his results indicate that this boundary layer
temperature would be far too low to heat the white dwarf deeply enough
to explain the observed cooling timescale. However, other boundary
layer solutions for different input parameters can reach substantially
higher temperatures; for instance, the solutions with $\md = 10^{-7}
\msyr$ reach $T_{BL} \sim 800,000$ K, and those with $\ms = 1 \msun$
reach $T_{BL} \sim 600,000$ K (Popham \& Narayan 1995). A solution
with increased $\md$ and $\ms$ would reach even higher temperatures.
This suggests that direct radiative heating across the disk--star
interface is likely to be most important in dwarf novae which contain
massive white dwarfs, or which have very high accretion rates during
outbursts.

Advective heating of the central star results from the deposition of
heated material onto the star. The most detailed treatment of this
process to date remains that of Prialnik \& Livio (1985), who
calculated the effects of accretion onto a 0.2 $\msun$ main-sequence
star. They varied the mass accretion rate and the fraction of the
accretion luminosity retained by the accreting material. FU Orionis
objects have high mass accretion rates $\dot M \simeq 10^{-4} \msyr$,
and the accreting material retains a large fraction $\alpha \simeq
0.375$ of the accretion luminosity, as discussed in \S 4. For these
values of $\dot M$ and $\alpha$, the $0.2 \msun$ main-sequence star
studied by Prialnik \& Livio undergoes dynamically unstable expansion.
If these results can be applied to the accreting stars in FU Orionis
objects, they suggest that advective heating will cause the star to
expand rapidly udring outbursts.  This may account for the rather
large radii inferred by disk models for the stars in FU Orionis
objects, as compared to the radii of T Tauri stars (e.g., Popham \et
1996). In addition, the advected energy added to the star during FU
Orionis outbursts may change the position of the stellar birthline in
the H-R diagram, as emphasized by Hartmann, Cassen, \& Kenyon (1996).

\subsection{Problems with One-Dimensional, Steady-State Models}

The heating of an accreting star by a luminous boundary layer is
difficult to study with one-dimensional, steady-state models. The
fundamental reasons for this are, first, although the mass accretion
rate may be constant throughout the disk, when the accreting material
reaches the star, it should build up on the stellar surface. To
accurately study this requires a model where $\md$ can vary with
radius, which must be time-dependent. Second, the approximations used
to treat the vertical structure of the disk must break down within the
star. 

Another problem with one-dimensional, steady-state boundary layer
models is the difficulty in assigning the location of the disk--star
boundary. Our model treats the disk, the boundary layer, and the
``star'' (the settling region which serves as an approximate version
of the outer layers of the star) using the same set of equations. We
have found that it can be quite difficult to tell where the disk ends
and the star begins. In fact, in our previous papers, we have used
several different criteria to define the disk--star boundary, where $R
= \rs$. In our T Tauri solution (Popham \et 1993), we assigned $\rs$
to be the point where the angular velocity $\O$ drops to half of its
peak value. In our CV solutions (Popham \& Narayan 1995), we chose the
point where the rate of energy dissipation by viscosity drops below
the rate of energy advection. Finally, in our FU Orionis solutions
(Popham 1996; Popham \et 1996), we chose the point where the radial
velocity of the accreting material drops below a set value, which we
took to be $1000 {\rm cm} \pers$.

In general, the ambiguity in the stellar radius is fairly small in
thin disk (CV and T Tauri) solutions. This is because the properties
of the disk vary over a short radial scale; however, this rapid
variation means that the small ambiguity in the location of $\rs$
results in fairly large uncertainties in the values of quantities such
as $T(\rs)$.  In thick disk solutions like the ones for FU Orionis
objects, the uncertainty in $\rs$ is much larger, because the
variations in the disk occur over quite large lengthscales, comparable
to $\rs$ itself.  Of course, the slower variations of the disk
properties keep the uncertainties in these quantities from becoming
too large.

The ambiguity in the location of the stellar radius affects the values
derived in this paper for the rates of stellar heating. For instance,
if we selected a smaller stellar radius for our CV and T Tauri
solutions, we would find a somewhat larger rate of heating by
advection, since the temperature of the accreting material rises
rapidly with decreasing $R$ in this region. In general, although the
uncertainty in the stellar radius may alter the heating rates we have
computed here, it should not affect the qualitative conclusions we
have drawn about which types of stellar heating will be important in
various types of accreting systems.

Two-dimensional, time-dependent models of boundary layers which
include the outer layers of the star should be able to resolve the
uncertainties in the location of the stellar radius and the radiative
flux across the disk--star interface. Such models have been
constructed by Kley for boundary layers in CVs (Kley 1989, 1991) and
pre-main-sequence stars (Kley \& Lin 1996).  However, thus far none of
these models has been run long enough to reach a steady state;
however, the most recent models of pre-main-sequence stars have been
carried out for as long as 1000 orbital periods (Kley \& Lin
1996). The initial conditions for these models take the temperature
distribution in the star to be that of a star undisturbed by
accretion. Thus initially there is a minimum in the temperature at
$\rs$, similar to the temperature minimum assumed in the steady-state
``cool'' boundary layer models discussed in \S 3. This region should
be heated by radiative diffusion from both sides, filling in the
temperature minimum. Kley \& Lin (1996) estimate that this should take
place in $\sim 10^5$ orbital periods, or about 100 years.
Unfortunately this timescale is still too long for the heating of this
region to be verified by current models. Nonetheless, if future models
of this type can be extended over substantially longer timescales,
they should be able to resolve this issue.

\acknowledgements 

This work was supported by NASA grant NAG5-2837 at the Center for
Astrophysics and grants NASA NAGW-1583, NSF AST 93-15133, and NSF PHY
91-00283 at the University of Illinois.

\newpage

\figcaption{(a) The effective temperature of the boundary layer and
inner disk, for a cataclysmic variable (solid line), an FU Orionis
object (short-dashed line), and a T Tauri star (long-dashed line). The
thin disk (CV and T Tauri) solutions have strong peaks in the
effective temperature due to the boundary layer, while the thick disk
FU Orionis solution has the boundary layer luminosity spread out over
a large region. (b) The fraction of the flux emitted at a given disk
radius which is intercepted by the stellar surface.}

\figcaption{(a) The incident flux on the stellar surface, in units of
$F_* = 3 G M_* \dot M / 8 \pi R_*^3$, for disks with a $T(r) \propto
r^{-3/4}$ temperature distribution (solid line) and a Shakura-Sunyaev
$T(r) \propto r^{-3/4} (1 - r^{-1/2})^{1/4}$ distribution (dashed
line). (b) The boundary layer and disk flux incident on the stellar
surface given by boundary layer solutions for a cataclysmic variable
(solid line), an FU Orionis object (long-dashed line), and a T Tauri
star (short-dashed line). The dotted lines show the incident flux that
would result from a Shakura-Sunyaev-type solution with the same
parameters. For the thin disk (CV and T Tauri) solutions, the flux
from the boundary layer produces a pronounced peak in the incident
flux on the stellar surface, but this peak only extends to 15 or 20
degrees above the disk plane. For the thick disk FU Orionis solution,
the boundary layer flux is spread more uniformly over most of the
stellar surface.}

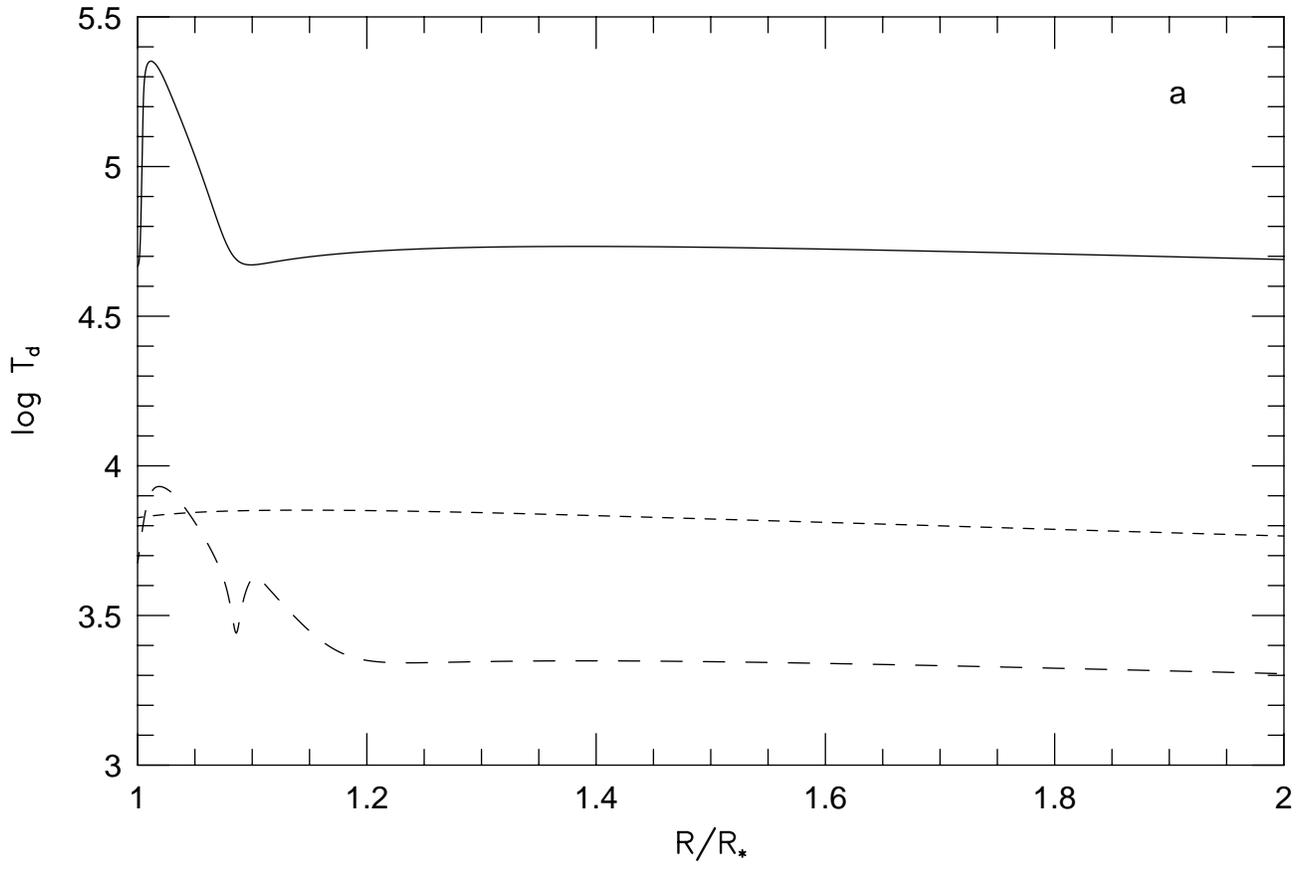
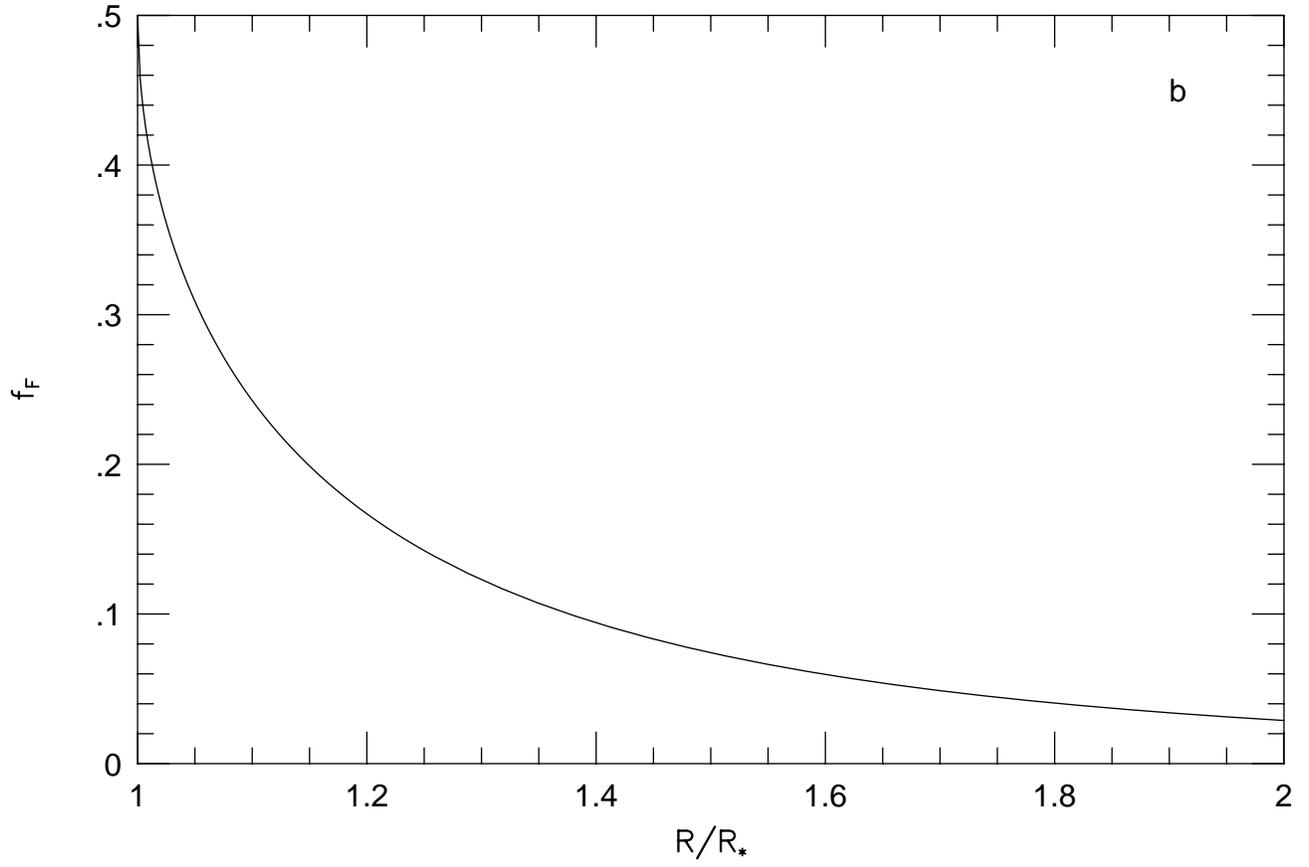
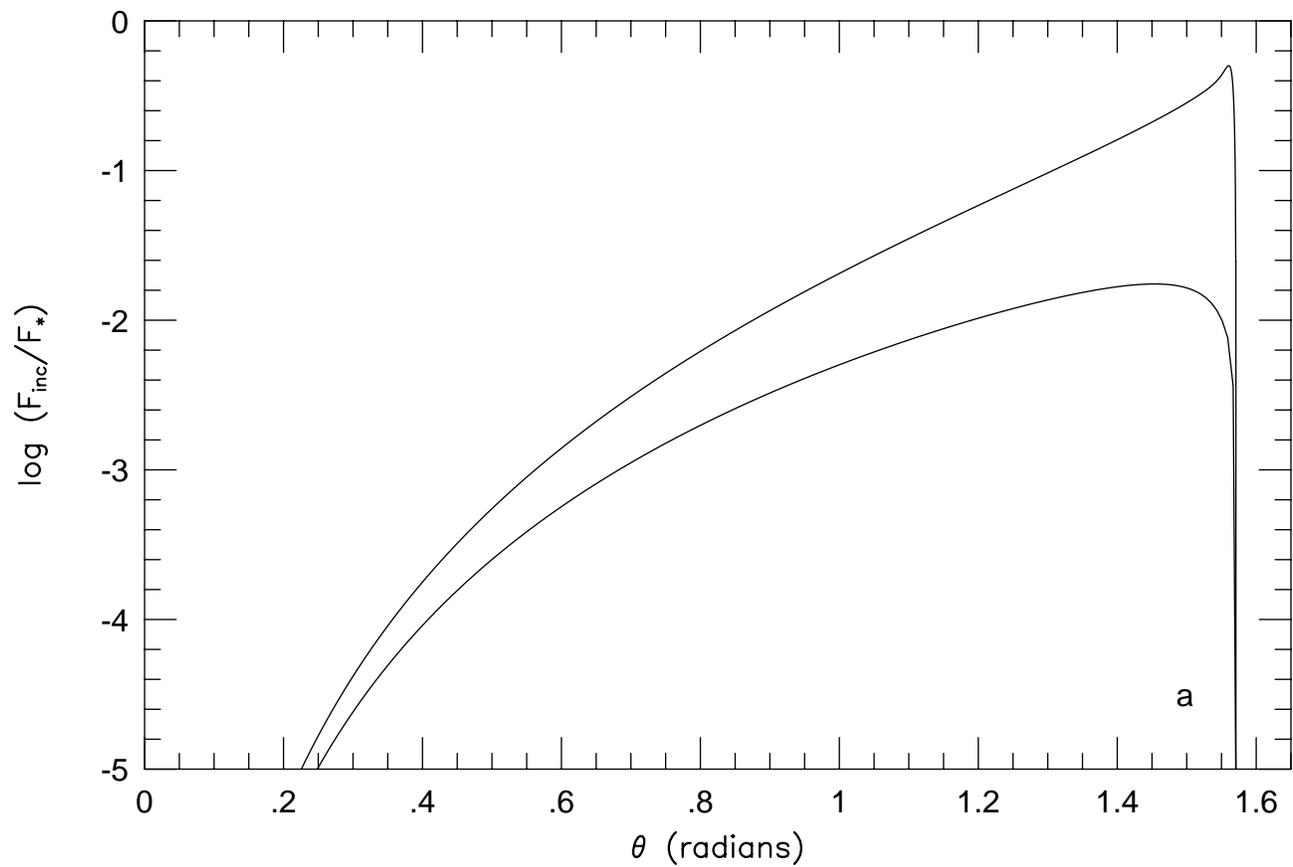
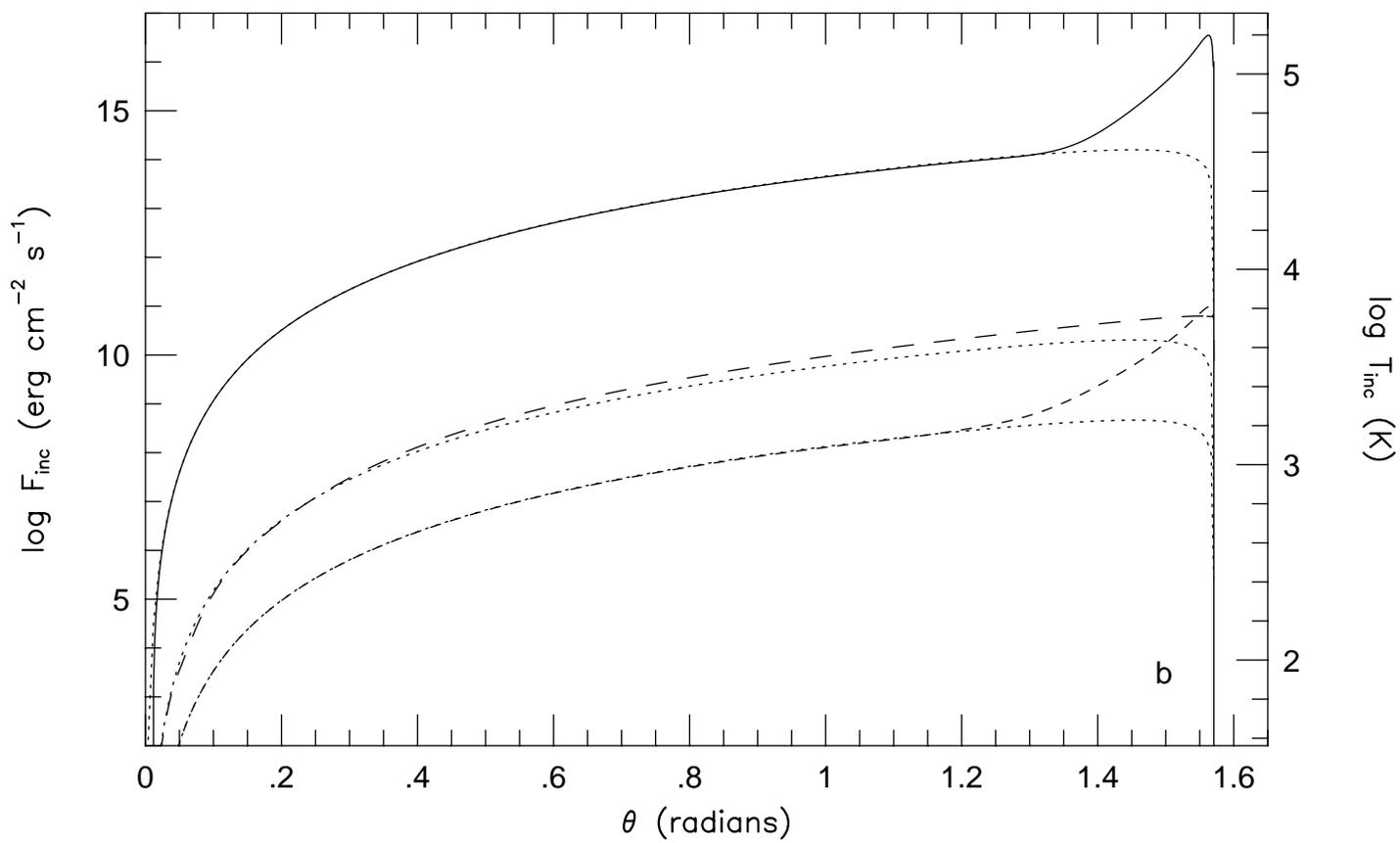
\end{document}